\documentclass[ pra,reprint,superscriptaddress,amsmath,amssymb,aps,longbibliography]{revtex4-2}

\def \beq {\begin{equation}}
\def \eeq {\end{equation}}
\def \ba {\begin{align}}
\def \ea {\end{align}}

\usepackage{epsfig}
\usepackage{amssymb}
\usepackage{amsfonts}
\usepackage{mathrsfs}
\usepackage{bm}
\usepackage{bbm,mathbbol}
\usepackage{braket,bigints}
\usepackage{stmaryrd,mathtools}
\usepackage{graphicx}
\usepackage{dcolumn}
\usepackage{xcolor}
\usepackage{bbold}
\usepackage{xparse}
\usepackage{siunitx}
\usepackage{braket,bigints}
\usepackage[T1]{fontenc}
\usepackage[
colorlinks=true,
linkcolor=blue,
citecolor=blue,
urlcolor=blue
]{hyperref}

\usepackage{cleveref}

\newcommand{\Tr}{\mathrm{Tr}}

\newcommand{\Rse}{R_{\textrm{se}}}
\newcommand{\Rop}{R_{\textrm{op}}}
\newcommand{\Rsd}{R_{\textrm{sd}}}

\renewcommand\bra[1]{{\langle{#1}|}}
\makeatletter
\renewcommand\ket[1]{%
\@ifnextchar\bra{\k@t{#1}\!}{\k@t{#1}}%
}
\newcommand\k@t[1]{{|{#1}\rangle}}
\makeatother

\usepackage{color}
\definecolor{mygreen}{rgb}{0,0.5,0}
\definecolor{mygrey}{rgb}{0.5,0.5,0.5}
\definecolor{myred}{rgb}{0.75,0,0}
\definecolor{myblue}{rgb}{0,0,0.75}
\definecolor{mymagenta}{cmyk}{0,1,0,0.12}
\definecolor{mycyan}{cmyk}{1,0,0,0.12}
\definecolor{myorange}{rgb}{1.,0.5,0}
\definecolor{myviolet}{rgb}{0.6,0.15,0.6}
\definecolor{mybrown}{cmyk}{0,0.50,1,0.41}

\usepackage{ulem}

\begin{document}
\title{Nuclear slowing-down factors in alkali-metal vapors}

\author{Vasiliki Koutrouli}
\affiliation{
Institute of Electronic Structure and Laser, Foundation for Research and Technology, 71110 Heraklion, Greece}

\author{Georgios Vasilakis}
\email{gvasilak@iesl.forth.gr}
\affiliation{
Institute of Electronic Structure and Laser, Foundation for Research and Technology, 71110 Heraklion, Greece}

\author{Kostas Mouloudakis}
\email{mouloudakisk@gmail.com}
\affiliation{Department of Physics, Princeton University, Princeton, New Jersey 08544, USA}

\begin{abstract}
Nuclear slowing-down factors account for the sharing of angular momentum between the electron and the nucleus in the effective Bloch description of alkali-metal spin dynamics in the spin-exchange-relaxation-free (SERF) regime. For collinear optical pumping and magnetic field, we find that longitudinal and transverse spin dynamics are characterized by different polarization-dependent slowing-down factors. Transverse dynamics are governed by the conventional factor $q(p)$, with $p$ the electron spin polarization, whereas longitudinal relaxation is governed by $q(p)+p\,dq(p)/dp$. The two factors coincide at zero polarization but differ substantially at high polarization, where the single-factor description overestimates the underlying pumping and relaxation rates inferred from longitudinal transients by factors approaching two to four, depending on the nuclear spin. We further derive closed-form expressions, valid at arbitrary polarization, for the residual spin-exchange relaxation of the transverse spin components at finite magnetic field along the pumping axis. This contribution is quadratic in the magnetic field and can become comparable to the zero-field linewidth at fields well within the SERF regime. The results are obtained by perturbation theory on the density-matrix dynamics linearized around the stationary state and are verified by numerical solutions of the microscopic density-matrix equation. These findings refine the effective Bloch description of alkali-metal spin dynamics and are directly relevant to atomic magnetometers and alkali-metal--noble-gas comagnetometers, including those used in precision searches for physics beyond the Standard Model.
\end{abstract}

\maketitle

\section{Introduction}

Optically polarized alkali-metal vapors are widely used in precision quantum sensing. Their collective spin dynamics underlie highly sensitive atomic magnetometers \cite{Budker2007,Kominis2003,PhysRevApplied.21.014023}, as well as alkali-metal--noble-gas comagnetometers operating as gyroscopes \cite{Kornack2005,Hedges2025,PhysRevApplied.19.044092} and as probes of physics beyond the Standard Model \cite{Vasilakis2009,Terrano2022,wqqq-s2bz}. In these systems, the measured signals are determined by the precession, optical pumping, and relaxation of the electronic spin polarization. An accurate effective description of these processes is therefore essential for quantitatively interpreting experimental signals and extracting the underlying physical information.

Particularly high sensitivities are achieved by magnetometers operating in the spin-exchange-relaxation-free (SERF) regime \cite{HapperTang1973,HapperTam1977,Allred2002,Kominis2003}. At the high alkali-metal densities of SERF magnetometers, spin-exchange collisions occur much faster than the Larmor precession and average the evolution of the two ground-state hyperfine manifolds into a single long-lived collective spin mode, whose spin-exchange relaxation vanishes in the zero-field limit.

Under high-buffer-gas-pressure conditions considered here, where the ground-state hyperfine structure is not optically resolved \cite{Appelt1998}, the spin dynamics are commonly described by effective Bloch equations for the electronic spin-polarization vector $\mathbf P = 2\langle\mathbf S\rangle$. A widely used form is \cite{ShahRomalis,Savukov2005,Ledbetter2008SERF,PhysRevApplied.21.014023,Padniuk2022,PhysRevA.107.043110,Hedges2025,TransientOE2021}:
\begin{equation}
\frac{d\mathbf P}{dt}
=\frac{1}{q(p)}
\left[
\gamma_e\,\mathbf B \times\mathbf P
+\Rop\,(\mathbf s_{\rm ph}-\mathbf P)
-\Rsd\,\mathbf P
\right].
\label{eq:BlochSERFstandard}
\end{equation}
Here $\Rop$ is the optical-pumping rate, $\mathbf s_{\rm ph}$ is the photon-spin polarization of the pump light (D\textsubscript{1} pumping is assumed throughout), $\gamma_e=1.761 \times 10^{11}\text{ s}^{-1}\text{T}^{-1}$ is the electron gyromagnetic ratio, and $\mathbf B$ is the magnetic field. The rate $\Rsd$ represents electron-spin relaxation mechanisms other than optical pumping and spin exchange, including spin-destruction collisions and probe-induced relaxation. Atomic motion is omitted here for simplicity. The numerical factor $q(p)$, with $p=|\mathbf P|$ the degree of electron spin polarization, is termed the nuclear slowing-down factor, as it reduces the precession and relaxation rates relative to those of an isolated electron spin. Its physical significance stems from the sharing of the atomic spin between the electron and the nucleus. The factor $q(p)$ was introduced in Ref.~\cite{Appelt1998} through the relation $\langle F_z\rangle=q(p)\,\langle S_z\rangle$, where the $z$ axis is taken along the optical-pumping direction and $F_z=I_z+S_z$. For small transverse excitations it applies to the transverse components as well, and in this form it was used in Ref.~\cite{Allred2002} to describe the transverse relaxation in the SERF regime and in Ref.~\cite{Savukov2005} to describe the slowing of the spin precession. For the spin-temperature distribution maintained by rapid spin exchange \cite{Appelt1998}, this proportionality yields the expressions listed in Table~\ref{tab:SlowingDownFactor0}, which are the slowing-down factors used throughout the literature.

\begin{table}[t]
\centering
\begin{tabular}{c c}
\hline\hline
$I$ & $q(p)$ \\
\hline
$\dfrac{3}{2}$ &
$\dfrac{6 + 2p^2}{1 + p^2}$ \\[8pt]
$\dfrac{5}{2}$ &
$\dfrac{38 + 52p^2 + 6p^4}{3 + 10p^2 + 3p^4}$ \\[8pt]
$\dfrac{7}{2}$ &
$\dfrac{22 + 70p^2 + 34p^4 + 2p^6}{1 + 7p^2 + 7p^4 + p^6}$ \\
\hline\hline
\end{tabular}
\caption{Nuclear slowing-down factors $q(p)$ for the alkali nuclear spins $I$, as functions of the electron spin polarization $p$
\cite{Savukov2005,Ledbetter2008SERF}.}
\label{tab:SlowingDownFactor0}
\end{table}

In writing Eq.~\eqref{eq:BlochSERFstandard}, it is implicitly assumed that the same polarization-dependent factor $q(p)$ governs Larmor precession, optical pumping, and relaxation of all components of the electronic polarization. However, longitudinal dynamics differ from transverse dynamics. A transverse perturbation rotates the spin distribution as a whole, tilting $\mathbf P$ while preserving the magnitude of the polarization. A longitudinal perturbation changes the magnitude of $\mathbf{P}$. Only in the unpolarized limit, where the stationary state is isotropic, does rotational symmetry force the transverse and longitudinal relaxation rates to coincide. Consequently, and contrary to the assumption built into Eq.~\eqref{eq:BlochSERFstandard}, longitudinal and transverse dynamics are slowed by different factors at finite polarization. Moreover, although Eq.~\eqref{eq:BlochSERFstandard} is written for arbitrary magnetic-field and optical-pumping directions, the slowing-down factors of Table~\ref{tab:SlowingDownFactor0} apply only when the static magnetic field is parallel to the optical-pumping axis.

An additional limitation of Eq.~\eqref{eq:BlochSERFstandard} concerns the residual effect of spin exchange at a small but finite static magnetic field, where a contribution quadratic in the field remains \cite{HapperTam1977,Savukov2005}. To our knowledge, an analytical expression for the coefficient of this relaxation term at arbitrary polarization has not previously been obtained. This contribution can become comparable to the zero-field transverse linewidth already at fields much smaller than the SERF scale $\Rse/\gamma_e$, with $\Rse$ the spin-exchange rate.

In this work, using perturbation theory on the linearized alkali-spin dynamics, we show that the transverse relaxation is governed by the conventional factor $q_T(p)=q(p)$ of Table~\ref{tab:SlowingDownFactor0}, whereas the longitudinal relaxation is governed by the distinct factor $q_L(p)=q(p)+p\,dq(p)/dp$. We further derive analytical expressions, valid at arbitrary spin polarization, for the residual finite-field spin-exchange relaxation of the transverse SERF mode. The analytical results are verified by numerical solutions of the density-matrix equation. Related eigenmode analyses at low field and low polarization can be found in Refs.~\cite{KatzFirstenberg2018,Xiao2021,PhysRevA.109.L040802,Dikopoltsev_2025,Tang2025}.

The paper is organized as follows. Section~\ref{sec:theory} derives the longitudinal and transverse slowing-down factors from perturbation theory on the linearized dynamics. Section~\ref{sec:finite_SE} analyzes the finite-field spin-exchange relaxation and the second-order corrections to the transverse dynamics. Section~\ref{sec:Numerical_verification} presents the numerical verification, and Section~\ref{sec:Conclusion} concludes.

\section{Longitudinal and transverse slowing-down factors}
\label{sec:theory}

We follow the formalism developed in Refs.~\cite{Appelt1998,MouloudakisVasilakis2026} and linearize the density-matrix equation around the stationary density matrix $\rho_{\rm ss}$. The density matrix is written as $\rho=\rho_{\rm ss}+\delta\rho$, with $\delta\rho$ a small deviation from the stationary state. The deviation is expanded in a basis of operators, and the expansion coefficients are collected in a vector $\mathbf x$. The linearized dynamics then take the form $\dot{\mathbf x}=\mathcal{A}\mathbf x$, with the drift matrix
\begin{equation}
\mathcal{A}=\mathcal{A}^{(B)}+\mathcal{A}^{(\text{se})} +\mathcal{A}^{(\text{sd})} +\mathcal{A}^{(\text{op})}, 
\label{eq:driftmatrix}
\end{equation}
where $\mathcal{A}^{(B)}$ is associated with the dc magnetic field, while $\mathcal{A}^{(\text{se})}$, $\mathcal{A}^{(\text{sd})}$ and $\mathcal{A}^{(\text{op})}$ are associated with spin-exchange, spin destruction and optical pumping processes, respectively. Explicit forms of these matrices are presented in \cite{MouloudakisVasilakis2026}. The observable relaxation rates are determined by the eigenvalues of the total drift matrix $\mathcal{A}$ (see supplementary material of \cite{MouloudakisVasilakis2026}). 

The dynamics described by the matrix $\mathcal{A}$ depend on the relative geometry of the optical pumping and the magnetic field. Here, we consider an axially symmetric configuration, commonly employed in experiments, in which the pump beam and the dc magnetic field $B_z$ are aligned along a common longitudinal axis, taken to be the $z$ axis. We take the pump helicity such that the stationary polarization is along $+z$, so that $p=P_z=2\langle S_z\rangle$. The eigenvalues of the drift matrix, and hence the physical relaxation rates and precession frequencies, are independent of the basis used to represent the density matrix. For this collinear geometry, the spherical-tensor basis \cite{HapperReview1972,MouloudakisVasilakis2026}, in which $\mathbf x$ collects the coefficients of the irreducible tensor operators $T^L_M(FF')$, is particularly convenient. Each tensor component $T^L_M(FF')$ has a definite azimuthal quantum number $M$ and acquires only a phase under rotations about the $z$ axis. Since the dynamics are invariant under such rotations, components with different $M$ do not couple, and the drift matrix is block-diagonal in $M$. 
Hyperfine coherences between the two ground-state manifolds ($F\neq F'$) oscillate at the ground-state hyperfine frequency, which is much larger than $\Rse$ and the other dynamical rates considered here. Their coupling to the slow $F=F'$ Zeeman modes therefore averages to zero to leading order, and we omit them in what follows (hyperfine-secular approximation) \cite{Appelt1998}.

The longitudinal polarization $P_z$ lies in the $M=0$ block, whereas the circular transverse components $P_\pm=P_x\pm iP_y$ lie in the $M=\pm1$ blocks. Thus, the longitudinal and transverse dynamics can be analyzed independently. The $M=+1$ and $M=-1$ blocks are related by complex conjugation, as required by the Hermiticity of the density matrix, and their spectra are therefore complex conjugates of one another. It is consequently sufficient to analyze a single transverse block.

Each $M$ block generally contains several eigenmodes. Writing an eigenvalue as $\lambda_n=-\Gamma_n+i\omega_n$, the real part determines the decay rate $\Gamma_n=-\operatorname{Re}\lambda_n$, while the imaginary part determines the precession frequency $\omega_n=\operatorname{Im}\lambda_n$. The spin dynamics are therefore generally a superposition of exponentially decaying and oscillating eigenmodes.

In the SERF regime, spin exchange defines the dominant timescale: $\gamma_e B_z,\; \Rop,\; \Rsd \ll \Rse$. The slow eigenvalues may then be obtained either by calculating the full spectrum of $\mathcal{A}$ and subsequently expanding in the small parameters $\Rop/\Rse$, $\Rsd/\Rse$, and $\gamma_e B_z/\Rse$, or directly by perturbation theory. The latter approach is more suitable for obtaining analytical expressions and will be used in what follows. This hierarchy also determines the partition of the drift matrix into unperturbed and perturbative contributions. We take $\mathcal{A}^{(\mathrm{se})}$ as the unperturbed matrix and $\delta\mathcal{A}=\mathcal{A}^{(\mathrm{sd})} +\mathcal{A}^{(\mathrm{op})}+\mathcal{A}^{(B)}$ as the perturbation. 

Spin-exchange collisions conserve both the number of atoms and the total angular momentum $\langle\mathbf F\rangle$ \cite{HapperReview1972, Appelt1998}. The null space of $\mathcal{A}^{(\mathrm{se})}$ is therefore four-dimensional, with one zero eigenvalue for each conserved quantity: the total population and the three components of $\langle\mathbf F\rangle$. All other eigenmodes decay on a timescale of order $1/\Rse$, as spin exchange drives the system toward the spin-temperature distribution compatible with the conserved quantities. The slow eigenvalues of the full dynamics develop from these zero modes under the perturbation, and the slowing-down factors follow by identifying the polarization-dependent renormalization of the Larmor frequency $\gamma_e B_z$ and of the rates $\Rsd$ and $\Rop$.

Under the axial symmetry described above, the four null modes are distributed among the $M$ blocks as follows. The $M=0$ block contains two of them, associated with conservation of the total population and of $\langle F_z\rangle$, whereas each of the $M=\pm1$ blocks contains a single null mode, associated with conservation of the corresponding transverse component $\langle F_{\pm}\rangle=\langle F_x\rangle\pm i\langle F_y\rangle$ of the total angular momentum. After the fast spin-exchange transients have decayed, the dynamics are governed by these slow modes. The perturbation $\delta\mathcal{A}$ partially lifts the zero-eigenvalue degeneracy and weakly admixes the slow modes with modes that decay at rates of order $\Rse$. The magnitude of this admixture is generically controlled by the small ratios $\gamma_e B_z/\Rse$, $\Rop/\Rse$, and $\Rsd/\Rse$.

We first consider the transverse blocks. Because the zero mode in each of the $M=\pm1$ blocks is nondegenerate, ordinary perturbation theory applies. The slow eigenvalues of the two blocks form a complex-conjugate pair; for $B_z>0$ we denote by $\lambda_T$ the member with $\operatorname{Im}\lambda_T\leq 0$, for which the second-order results of Sec.~\ref{sec:finite_SE} take their simplest form. To first order, the slow eigenvalue contains no explicit dependence on $\Rse$. Its real part gives the transverse decay rate
\begin{equation}
\frac{1}{T_2}=\frac{\Rsd+\Rop}{q_T(p)},
\label{eq:GammaT}
\end{equation}
while its imaginary part gives the effective spin-precession frequency
\begin{equation}
\omega_T=-\operatorname{Im}\lambda_T=\frac{\gamma_e B_z}{q_T(p)}.
\label{eq:omegaT}
\end{equation}
The resulting transverse slowing-down factor is the conventional factor $q_T(p)=q(p)$, listed in Table~\ref{tab:SlowingDownFactor0}. These transverse results follow from the linearized description and therefore hold for small deviations of the transverse polarization from the axial stationary state.

We next consider the longitudinal $M=0$ block. By contrast, its zero eigenvalue is twofold degenerate, and degenerate perturbation theory is therefore required. Within the linear-Zeeman and hyperfine-secular approximation adopted here, the coupling to a longitudinal magnetic field is proportional to $M$ and consequently vanishes in the $M=0$ block. The relevant first-order perturbation is $\mathcal{A}^{(\mathrm{sd})}+\mathcal{A}^{(\mathrm{op})}$. Diagonalizing this perturbation within the two-dimensional null space yields two eigenvalues. One remains exactly zero, because every process in the dynamics conserves the total population. This mode carries no physical excitation, since any physical perturbation conserves the population, and merely expresses that stationary solutions exist for any total number of atoms. The second eigenvalue describes longitudinal relaxation and takes the form
\begin{equation}
\lambda_L=-\frac{\Rsd+\Rop}{q_L(p)}.
\label{eq:lambdaL}
\end{equation}
The resulting longitudinal slowing-down factor can be written in the compact form
\begin{equation}
q_L(p)=q(p)+p\frac{dq(p)}{dp},
\label{eq:qL_general}
\end{equation}
with $q(p)$ the conventional transverse factor of Table~\ref{tab:SlowingDownFactor0}. Explicit expressions for the nuclear-spin values considered here are listed in Table~\ref{tab:SlowingDownFactorLongitudinal}.

\begin{table}[t]
\centering
\begin{tabular}{c c}
\hline\hline
$I$ & $q_L(p)$ \\
\hline
$\dfrac{3}{2}$ &
$\dfrac{2 \left(p^4+3\right)}{\left(p^2+1\right)^2}$ \\[8pt]
$\dfrac{5}{2}$ &
$\dfrac{2 \left(9 p^8+12 p^6+134 p^4+44 p^2+57\right)}{\left(3 p^4+10 p^2+3\right)^2}$ \\[8pt]
$\dfrac{7}{2}$ &
$\dfrac{2 \left(p^{12}+4 p^{10}+49 p^8+64 p^6+99 p^4+28 p^2+11\right)}{\left(p^6+7 p^4+7 p^2+1\right)^2}$ \\
\hline\hline
\end{tabular}
\caption{Longitudinal nuclear slowing-down factors $q_L(p)$ for
different nuclear-spin values $I$.}
\label{tab:SlowingDownFactorLongitudinal}
\end{table}

The same relation follows directly from the treatment of Ref.~\cite{Appelt1998}. Their paramagnetic coefficient, defined in Eq.~(118) of Ref.~\cite{Appelt1998}, satisfies $1+\epsilon(I,p)=\langle F_z\rangle/\langle S_z\rangle=q(p)$. Restricting their Eq.~(124) to the spin-destruction and optical-pumping processes considered here gives
\begin{equation}
\frac{d}{dt}\!\left[q(p)\,\langle S_z\rangle\right] =-\left(\Rsd+\Rop\right)\langle S_z\rangle+\frac{\Rop s_{\rm ph}}{2},
\label{eq:Appelt124}
\end{equation}
where $s_{\rm ph}=\mathbf s_{\rm ph}\cdot\hat{\mathbf z}$. Using $p=2\langle S_z\rangle$, Eq.~\eqref{eq:Appelt124} becomes
\begin{equation}
\left[ q(p)+p\frac{dq(p)}{dp} \right]\,\frac{dp}{dt} =-\left(\Rsd+\Rop\right)p+\Rop s_{\rm ph} ,
\label{eq:longitudinal_relaxation}
\end{equation}
reproducing Eq.~\eqref{eq:qL_general}.

This comparison shows that the polarization-dependent longitudinal slowing-down factor is already implicit in Eq.~(124) of Ref.~\cite{Appelt1998}. However, the difference between the longitudinal and transverse factors is not represented in the single-factor effective Bloch equation, Eq.~\eqref{eq:BlochSERFstandard}. The present work makes this distinction explicit and establishes the corresponding form of the effective dynamics.

Unlike the transverse result, Eq.~\eqref{eq:longitudinal_relaxation} for the longitudinal evolution is not restricted to small deviations from the stationary state. Rapid spin exchange keeps the density matrix close to the spin-temperature manifold, on which $\langle F_z\rangle=q(p)\,p/2$, and the chain rule then yields the nonlinear longitudinal evolution without linearization, with $q_L$ evaluated at the instantaneous polarization, up to corrections of higher order in the ratios of the slow rates to $\Rse$.

\begin{figure}[t]
\centering
\includegraphics[width=\columnwidth]{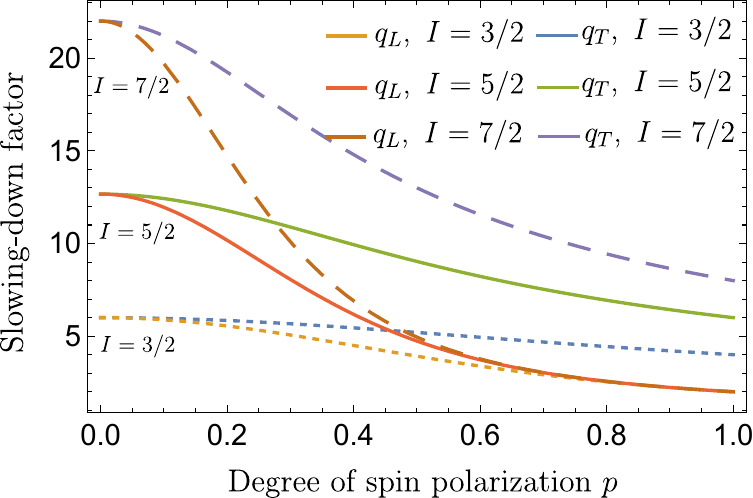}
\caption{Polarization dependence of the longitudinal and transverse nuclear slowing-down factors, $q_L(p)$ and $q_T(p)$, for nuclear spins $I=3/2$, $5/2$, and $7/2$. The two slowing-down factors coincide in the unpolarized limit ($p \rightarrow 0$) but differ increasingly at finite polarization, reflecting the distinction between changes in the magnitude and orientation of the spin polarization.}
\label{fig:qLT}
\end{figure}

The two slowing-down factors, $q_T(p)$ and $q_L(p)$, coincide in the unpolarized limit but differ increasingly at finite polarization, as illustrated in Fig.~\ref{fig:qLT}. At full polarization the two factors reach $q_T(1)=2I+1$ and $q_L(1)=2$. Longitudinal relaxation at high polarization therefore proceeds faster than Eq.~\eqref{eq:BlochSERFstandard} predicts by a factor approaching $(2I+1)/2$, and rates extracted from longitudinal transients using the conventional single-factor description are overestimated by the same factor. Since the polarization itself decays during a relaxation transient, this discrepancy is largest at the early times of the decay and diminishes as $p$ decreases.

Both factors can be viewed as directional derivatives of the relation $\langle\mathbf F\rangle=q(p)\,\langle\mathbf S\rangle$ between the conserved total angular momentum and the measured electronic spin, which holds for a density matrix of spin-temperature form. A transverse perturbation leaves $p$ fixed, so that $\delta\langle F_{+}\rangle=q(p)\,\delta\langle S_{+}\rangle$ and $q_T=q(p)$; a longitudinal perturbation changes $p$, so that $\delta\langle F_z\rangle=\left[q(p)+p\,dq(p)/dp\right]\delta\langle S_z\rangle$, recovering $q_L$ of Eq.~\eqref{eq:qL_general}.

\section{Finite-field spin-exchange relaxation}
\label{sec:finite_SE}

The description derived so far is accurate to first order in the perturbations. The same perturbative framework yields the corrections of second order, to both the relaxation rates and the precession frequencies. The correction quadratic in the magnetic field is of particular relevance here. In many experiments the Zeeman rate $\gamma_e B_z$, although small compared with $\Rse$, can substantially exceed the optical-pumping and spin-relaxation rates. The correction quadratic in the field can then dominate the other second-order terms and can become comparable to the first-order linewidth, as observed already in early SERF experiments \cite{Allred2002}.

At finite longitudinal magnetic field, the Zeeman coupling of the slow transverse mode to the rapidly relaxing modes of the same $M$ block produces a second-order contribution to the real part of the transverse eigenvalue \cite{HapperTam1977}. Keeping the field-quadratic term and temporarily omitting the remaining second-order corrections, we obtain
\begin{equation}
\lambda_T =-\frac{\Rop+\Rsd}{q_T(p)}-C_T(p)\frac{(\gamma_e B_z)^2}{\Rse}
-i\frac{\gamma_e B_z}{q_T(p)},
\label{eq:lambdaT2}
\end{equation}
written for the transverse block analyzed above, with the conjugate block giving the complex-conjugate eigenvalue. Here, $C_T(p)$ is a dimensionless polarization-dependent coefficient. It follows from the second-order correction
\begin{equation}
\lambda_n^{(2)}
=\sum_{m\neq n}
\frac{\langle \ell_n^{(0)}| \mathcal{A}^{(B)}|r_m^{(0)}\rangle
      \langle \ell_m^{(0)}| \mathcal{A}^{(B)}|r_n^{(0)}\rangle}
     {\lambda_n^{(0)}-\lambda_m^{(0)}},
\label{eq:secondorderPT}
\end{equation}
where $|r_n^{(0)}\rangle$ and $\langle\ell_n^{(0)}|$ are the right and left eigenvectors of $\mathcal{A}^{(\mathrm{se})}$ (Appendix~\ref{sec:ApndxA: Perturbation_Theory}) and the sum runs over the fast spin-exchange modes, whose relaxation rates are of order $\Rse$. Since $\mathcal{A}^{(B)}\propto\gamma_e B_z$, the leading correction to the transverse relaxation scales as $(\gamma_e B_z)^2/\Rse$.

\begin{figure}[t]
\centering
\includegraphics[width=\columnwidth]{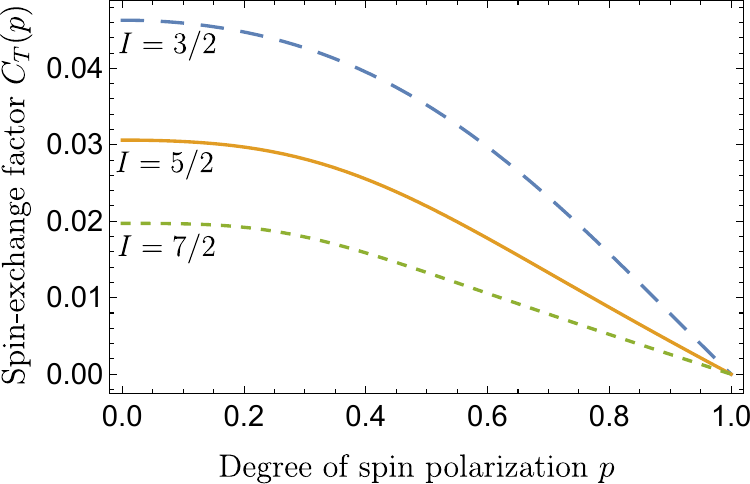}
\caption{Polarization dependence of the quadratic spin-exchange relaxation factor $C_T(p)$, for nuclear spins $I=3/2$, $5/2$, and $7/2$.}
\label{fig:SE_factor_C_T}
\end{figure}

Evaluating Eq.~\eqref{eq:secondorderPT} yields $C_T(p)$ in closed form. The resulting expressions for the alkali nuclear spins are collected in Table~\ref{tab:CT}, and their polarization dependence is shown in Fig.~\ref{fig:SE_factor_C_T}. In the unpolarized limit,
\begin{equation}
C_T(0)=\frac{q(0)^2-(2I+1)^2}{2\,q(0)^3},
\label{eq:CT0}
\end{equation}
which is the known low-polarization result \cite{HapperTam1977,Savukov2005}, with $C_T(0)=5/108,\;210/6859,\;105/5324$ for $I=3/2,\,5/2,\,7/2$. At full polarization, $C_T(1)=0$, and the finite-field broadening vanishes as $(1-p^2)$. This light narrowing is a consequence of the conservation of angular momentum in spin-exchange collisions: in a fully polarized vapor every atom occupies the stretched state, which carries the maximum total spin, and binary collisions that conserve the total spin cannot transfer atoms out of it.

At intermediate polarizations, $C_T(p)$ differs substantially from the substitution $\tilde C_T(p)\equiv[q(p)^2-(2I+1)^2]/[2q(p)^3]$ of the polarization-dependent slowing-down factor into Eq.~\eqref{eq:CT0}: the ratio $\tilde C_T(p)/C_T(p)$ increases monotonically from unity at $p=0$ and approaches $(6I-1)(2I+1)/[2(6I+1)]$ as $p\to1$, so that this substitution overestimates the residual broadening by factors approaching $8/5$, $21/8$, and $40/11$ for $I=3/2$, $5/2$, and $7/2$, respectively.

Including the field-quadratic correction, the leading relaxation rates of the slow modes are
\begin{equation}
\frac{1}{T_{1}}=\frac{\Rop+\Rsd}{q_L(p)},
\qquad
\frac{1}{T_{2}}=\frac{\Rop+\Rsd}{q_T(p)}
+C_T(p)\,\frac{(\gamma_e B_z)^2}{\Rse}.
\label{eq:T2eff}
\end{equation}
The transverse precession frequency remains that of Eq.~\eqref{eq:omegaT}.  Within the linear-Zeeman and hyperfine-secular approximation adopted here, a longitudinal field does not act in the $M=0$ block and therefore modifies neither the longitudinal rate nor the stationary polarization.

The field-dependent and zero-field contributions to $1/T_2$ are equal at the critical field
\begin{equation}
B_c(p)=\frac{1}{\gamma_e}\sqrt{\frac{\Rse(\Rop+\Rsd)}{C_T(p)\,q_T(p)}}
=B_{\rm SERF}\sqrt{\frac{\Rop+\Rsd}{C_T(p)\,q_T(p)\,\Rse}}\,,
\label{eq:Bc}
\end{equation}
where $B_{\rm SERF}=\Rse/\gamma_e$ is the conventional SERF field scale. Since $\Rop+\Rsd\ll\Rse$, $B_c$ can lie far below $B_{\rm SERF}$ over a broad range of spin polarization, so appreciable finite-field spin-exchange broadening may occur well within the conventional SERF condition. As $p\to1$, however, $C_T(p)\to0$ and $B_c$ increases because of light narrowing, and near full polarization it may approach or exceed $B_{\rm SERF}$. The perturbative treatment remains valid while $\gamma_e B_z\ll\Rse$.

\begin{table}[t]
\centering
\begin{tabular}{c c}
\hline\hline
$I$ & $C_T(p)$ \\
\hline
$\dfrac{3}{2}$ &
$\dfrac{(1-p^2)(1+p^2)\left(45+44p^2-9p^4\right)}
       {4\,(3+p^2)^3\,(9+7p^2)}$ \\[10pt]
$\dfrac{5}{2}$ &
$\dfrac{2(1-p^2)(3+p^2)(1+3p^2)\,\Pi_{5/2}(p)}
       {(19+26p^2+3p^4)^3\,(100+269p^2+72p^4)}$ \\[10pt]
$\dfrac{7}{2}$ &
$\dfrac{(1-p^2)(1+p^2)(1+6p^2+p^4)\,\Pi_{7/2}(p)}
       {4\,(11+35p^2+17p^4+p^6)^3\,\Sigma_{7/2}(p)}$ \\
\hline\hline
\end{tabular}
\caption{Polarization-dependent coefficients $C_T(p)$ of the residual spin-exchange relaxation $C_T(p)(\gamma_e B_z)^2/\Rse$, with $\Pi_{5/2}(p)=3500+13615p^2+7818p^4+6384p^6+1146p^8-207p^{10}$,
$\Pi_{7/2}(p)=10418625+95382630p^2+226669002p^4+351541614p^6+270069708p^8+80829458p^{10}+23781478p^{12}+1287450p^{14}-262845p^{16}$,
and $\Sigma_{7/2}(p)=99225+571041p^2+515203p^4+68931p^6$.}
\label{tab:CT}
\end{table}

The perturbative expansion can be carried to second order in all of the perturbations simultaneously. In addition to the field term of Eq.~\eqref{eq:lambdaT2}, the transverse relaxation then acquires corrections of order $\Rsd^2/\Rse$, $\Rop^2/\Rse$, and $\Rsd\Rop/\Rse$, and the precession frequency acquires corrections of order $\gamma_e B_z\Rsd/\Rse$ and $\gamma_e B_z\Rop/\Rse$. The corresponding coefficients are derived in Appendix~\ref{sec:ApndxB:second_order_correction} and listed in Table~\ref{tab:second_order_transverse}. 

At the stationary state of continuous circularly polarized D\textsubscript{1} pumping ($s_{\rm ph}=1$), where $p=\Rop/(\Rop+\Rsd)$, the six second-order contributions combine into a single term (Appendix~\ref{sec:ApndxB:second_order_correction}). Through second order in the perturbations, the transverse eigenvalue then takes the compact form
\begin{equation}
\lambda_T=-\frac{\zeta}{q_T(p)}+C_T(p)\,\frac{\zeta^{2}}{\Rse},
\qquad \zeta=(\Rop+\Rsd)+i\gamma_e B_z .
\label{eq:lambdaT_zeta}
\end{equation}
At fixed stationary polarization, the dependence of the transverse eigenvalue on the total relaxation rate $\Rop+\Rsd$ and on the longitudinal field thus combines into the single complex variable $\zeta$, extending the low-polarization structure of Ref.~\cite{HapperTam1977} to arbitrary polarization. The same function $C_T(p)$ governs the field-quadratic broadening, the second-order rate corrections, and the frequency shifts.

\begin{figure}[ht]
\centering
\includegraphics[width=\columnwidth]{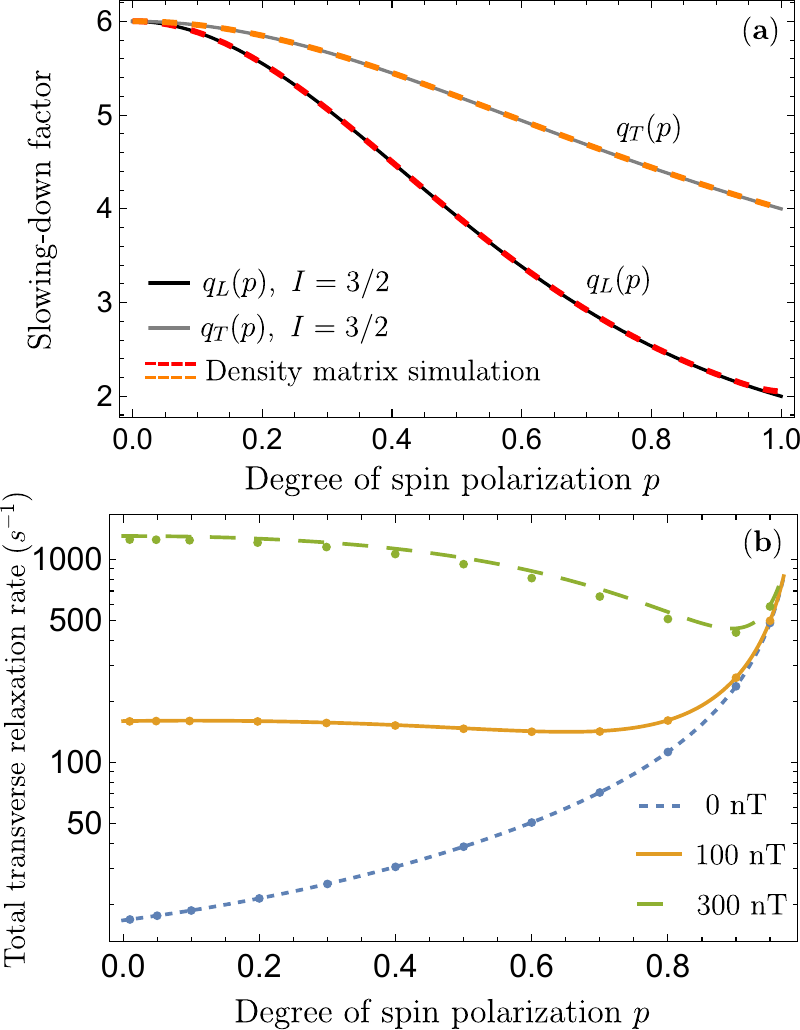}
\caption{
\textbf{(a)} Longitudinal and transverse slowing-down factors extracted from the slow eigenvalues of the linearized dynamics obtained from Eq.~\eqref{eq:density_matrix} (red and orange dashed curves), compared with the analytical expressions (black and gray solid curves), for $I=3/2$ and $\mathbf{B}=0$.
\textbf{(b)} Transverse relaxation rate as a function of the stationary spin polarization $p$. Colored points show the decay rate of the slow transverse eigenmode, while the solid curves show the analytical prediction of Eq.~\eqref{eq:lambdaT_zeta}, including all contributions through second order in the magnetic field, spin-destruction rate, and optical-pumping rate. The calculation is performed for $I=3/2$, with $\Rse=10^5~\si{\per\second}$ and $\Rsd=10^2~\si{\per\second}$. Circularly polarized D\textsubscript{1} pumping along the $z$ axis is assumed, with $s_{\rm ph}=1$ and $\Rop=p\,\Rsd/(1-p)$. Results are shown for longitudinal magnetic fields $B_z=0$, $100$, and $300~\si{\nano\tesla}$, as indicated in the legend.
}
\label{fig:DM_ver}
\end{figure}

\section{Numerical verification}
\label{sec:Numerical_verification}

We verify the analytical results against numerical solutions of the density-matrix equation, Eq.~\eqref{eq:density_matrix}, for a \textsuperscript{87}Rb vapor under high-buffer-gas-pressure optical pumping on the D\textsubscript{1} line (Appendix~\ref{sec:ApndxC: DM simulation}).

Equation~\eqref{eq:density_matrix} is nonlinear in the density matrix $\rho$ because the spin-exchange mean-field state depends on $\rho$ both through the reduced nuclear density matrix ${\rm Tr}_S[\rho]$ and through the mean electron spin $\langle\mathbf S\rangle=\mathrm{Tr}[\mathbf S\,\rho]$. The product of these two factors generates a term quadratic in $\rho$. The relaxation rates of the slow modes are therefore not obtained as eigenvalues of a fixed linear operator appearing in the original equation of motion. Instead, we first determine the stationary state $\rho_{\rm ss}$ and then linearize the full nonlinear dynamics around it by writing $\rho=\rho_{\rm ss}+\delta\rho$. To first order in $\delta\rho$,
\begin{equation}
\frac{d\,\delta\rho}{dt}=\mathcal{L}_{\rm lin}[\delta\rho],
\label{eq:linearized}
\end{equation}
where $\mathcal{L}_{\rm lin}$ is the Liouvillian obtained by linearizing the density-matrix equation about $\rho_{\rm ss}$ (see Appendix~\ref{sec:ApndxC: DM simulation} for details). For the numerical evaluation we work in Liouville space \cite{Appelt1998}, vectorizing $\delta\rho$ so that $\mathcal{L}_{\rm lin}$ is represented by an ordinary matrix whose eigenvalues are computed directly. For the axially symmetric geometry considered here, the linearized Liouvillian retains the block structure in $M$ discussed above. At zero magnetic field, the numerical slowing-down factors
\begin{equation}
q_L^{\rm num}=\frac{\Rsd+\Rop}{-\operatorname{Re}\lambda_L},
\qquad
q_T^{\rm num}=\frac{\Rsd+\Rop}{-\operatorname{Re}\lambda_T},
\end{equation}
obtained from the slow longitudinal and transverse eigenvalues of $\mathcal{L}_{\rm lin}$, reproduce the analytical expressions in Tables~\ref{tab:SlowingDownFactor0} and~\ref{tab:SlowingDownFactorLongitudinal}, as shown in Fig.~\ref{fig:DM_ver}(a). At finite longitudinal magnetic field, the real part of the slow transverse eigenvalue agrees with the complete second-order prediction of Eq.~\eqref{eq:lambdaT_zeta}, as shown in Fig.~\ref{fig:DM_ver}(b). Small deviations at the largest field considered, $B_z=300~\si{\nano\tesla}$, are consistent with higher-order magnetic-field corrections neglected in the second-order expansion. The imaginary part of $\lambda_T$ provides an independent verification of the frequency renormalization $\gamma_e B_z/q_T(p)$.

\begin{figure}[ht]
\centering
\includegraphics[width=\columnwidth]{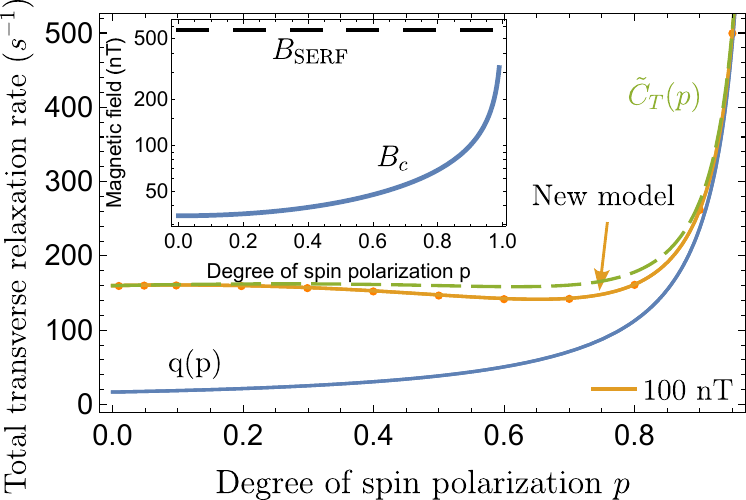}
\caption{Comparison between the conventional and corrected descriptions of the transverse relaxation rate as a function of spin polarization. The blue curve shows the conventional prediction, $(R_{\rm op}+R_{\rm sd})/q_T(p)$, while the orange curve includes the finite-field spin-exchange contribution $C_T(p)(\gamma_e B_z)^2/R_{\rm se}$. The green dashed curve instead includes the finite-field spin-exchange correction evaluated using $\tilde{C}_T(p)$. The calculation is shown for $B_z=100~\si{\nano\tesla}$, $R_{\rm se}=10^5~\si{\per\second}$, and $\Rsd=10^2~\si{\per\second}$, with circularly polarized D\textsubscript{1} pumping ($s_{\rm ph}=1$), for which $R_{\rm op}=p\,R_{\rm sd}/(1-p)$. The inset shows the polarization-dependent critical field $B_c(p)$, at which the quadratic spin-exchange broadening equals the zero-field transverse linewidth, together with the SERF field scale $B_{\rm SERF}$ (black dashed line). The separation between these two field scales demonstrates that appreciable spin-exchange broadening can occur while the vapor remains well within the SERF condition.
}
\label{fig:conventional_new}
\end{figure}

Figure~\ref{fig:conventional_new} illustrates the practical consequence of the finite-field correction. The conventional description, which includes only the slowing-down factor $q_T(p)$, substantially underestimates the transverse relaxation rate once the quadratic spin-exchange contribution becomes appreciable (blue curve). This occurs even though the applied field remains well below the SERF scale $B_{\rm SERF}$. In contrast, the model that naively includes $\tilde{C}_T(p)$ overestimates the transverse relaxation rate (green curve). The difference becomes even more significant at higher magnetic fields. At full polarization, $C_T(1)=0$ and the relaxation rate approaches $(R_{\rm op}+R_{\rm sd})/q_T(p)$, as shown in the figure.

\section{Conclusion}
\label{sec:Conclusion}
In conclusion, we have revisited the Bloch description of alkali-metal spin dynamics in the SERF regime by analyzing the slow eigenvalues of the microscopic drift matrix. Longitudinal and transverse spin dynamics are governed by different polarization-dependent slowing-down factors. The transverse factor is the conventional $q_T(p)=q(p)$, whereas the longitudinal factor is $q_L(p)=q(p)+p\,dq(p)/dp$. Therefore, pumping and relaxation rates inferred from longitudinal transients using the single-factor description can be overestimated. The difference between the longitudinal and transverse slowing-down factors indicates that geometries in which these components are coupled, such as those involving a magnetic field transverse to the pumping axis, generally require a modified effective description. We have further derived closed-form expressions, valid at arbitrary polarization, for the finite-field spin-exchange broadening of the transverse resonance, which can dominate the linewidth at fields well below the conventional SERF scale $\Rse/\gamma_e$. At the pumped stationary state, all second-order corrections combine into a single compact expression governed by one polarization-dependent function $C_T(p)$. All analytical results are confirmed by numerical solutions of the density-matrix equation. The results of this work are directly relevant to the quantitative modeling of alkali-metal spin dynamics in atomic magnetometers and in alkali-metal--noble-gas comagnetometers used for precision measurements and searches for physics beyond the Standard Model.

\appendix

\section{Perturbation theory}
\label{sec:ApndxA: Perturbation_Theory}

We consider a drift matrix $\mathcal{A} + \delta \mathcal{A}$, where $\mathcal{A}$ defines the unperturbed problem and $\delta \mathcal{A}$ is a small perturbation. We do not assume that $\mathcal{A}$ is Hermitian, but we assume that it is diagonalizable. Its right and left eigenvectors satisfy $\mathcal{A}\,|r_n^{(0)}\rangle = \lambda_n^{(0)} |r_n^{(0)}\rangle$ and $\langle \ell_n^{(0)}|\,\mathcal{A} =\lambda_n^{(0)} \langle \ell_n^{(0)}|$. Since $\mathcal{A}$ is generally non-Hermitian, its left and right eigenvectors need not be related by Hermitian conjugation. They may be chosen to satisfy the biorthonormality condition $\langle \ell_m^{(0)} | r_n^{(0)} \rangle =\delta_{mn}$, with the corresponding completeness relation $\sum_n |r_n^{(0)}\rangle\langle \ell_n^{(0)}|=\mathbb{1}$.

For a nondegenerate eigenvalue, the first-order correction induced by $\delta \mathcal{A}$ is $\lambda_n^{(1)} =\langle \ell_n^{(0)}| \delta \mathcal{A} |r_n^{(0)}\rangle$. The leading change in the eigenvalue of a dynamical mode is thus determined by the matrix element of the perturbation between the corresponding left and right eigenvectors. The first-order correction to the right eigenvector is
\begin{equation}
|r_n^{(1)}\rangle = \sum_{m\neq n}
\frac{ \langle \ell_m^{(0)}| \delta \mathcal{A}|r_n^{(0)}\rangle}
{\lambda_n^{(0)}  -\lambda_m^{(0)}}|r_m^{(0)}\rangle,
\end{equation}
where we choose the normalization $\langle \ell_n^{(0)}|r_n^{(1)}\rangle = 0$.

The second-order correction to a nondegenerate eigenvalue is
\begin{equation}
\lambda_n^{(2)}
=\sum_{m\neq n}
\frac{\langle \ell_n^{(0)}|\delta \mathcal{A}|r_m^{(0)}\rangle
      \langle \ell_m^{(0)}|\delta \mathcal{A}|r_n^{(0)}\rangle}
     {\lambda_n^{(0)}-\lambda_m^{(0)}}
=\langle \ell_n^{(0)}|\,\delta \mathcal{A}\,|r_n^{(1)}\rangle ,
\label{eq:Second_order_pert_eigen}
\end{equation}
where the second form expresses the second-order shift directly in terms of the first-order correction to the right eigenvector.

If an eigenvalue $\lambda^{(0)}$ is $g$-fold degenerate, first-order perturbation theory requires diagonalizing the perturbation within the degenerate subspace. Let $\{|r_\alpha^{(0)}\rangle\}_{\alpha=1}^g$ and $\{\langle \ell_\alpha^{(0)}|\}_{\alpha=1}^g$ span the corresponding right and left subspaces. The effective perturbation matrix is $W_{\alpha\beta} = \langle \ell_\alpha^{(0)} | \delta \mathcal{A} | r_\beta^{(0)} \rangle$, and the first-order eigenvalue shifts are $\lambda \simeq \lambda^{(0)} + \mu_j$, with $\mu_j \in \mathrm{eig}(W)$.

\subsection*{Resolved Zeeman resonances}

The perturbation theory developed above also applies in the hierarchy opposite to the SERF regime of the main text, in which the Zeeman splitting of the two hyperfine manifolds is large compared with all relaxation rates, $\omega_0=\gamma_e B_z/(2I+1)\gg\Rse,\Rop,\Rsd$.
The field is still assumed small compared with the hyperfine splitting, so that nonlinear-Zeeman corrections can be neglected. In this case, the magnetic-field matrix $\mathcal{A}^{(B)}$ defines the unperturbed problem, while spin exchange, spin destruction, and optical pumping are treated perturbatively. In a transverse $M=\pm1$ block, the two ground-state hyperfine manifolds have zeroth-order precession frequencies of opposite sign, $\lambda^{(0)}_{\pm}=\pm i\omega_0$, and the corresponding degeneracies are treated by degenerate perturbation theory. This is the resolved-resonance limit considered in Ref.~\cite{Appelt1998}. Spin exchange contains both matrix elements within a given zeroth-order resonant subspace and matrix elements coupling the two oppositely precessing hyperfine manifolds. The former contribute to the relaxation rates already at first order. The latter couple zeroth-order eigenvalues separated by approximately $2\omega_0$ and therefore affect the eigenvalues only at second order in $\Rse/\omega_0$. Thus, to first order in the resolved-resonance limit, the two hyperfine resonance blocks may be treated independently, while their mutual spin-exchange mixing produces higher-order corrections.

\section{Complete second-order correction to the transverse SERF eigenvalue}
\label{sec:ApndxB:second_order_correction}

Here we  apply the second-order expression of Eq.~\eqref{eq:Second_order_pert_eigen} to the slow transverse SERF mode. As in the main text, $\mathcal{A}^{(\rm se)}$ defines the unperturbed problem, while $\mathcal{A}^{(\rm sd)}$, $\mathcal{A}^{(\rm op)}$, and $\mathcal{A}^{(B)}$ are treated as perturbations. Retaining all pairwise combinations gives
\begin{align}
\lambda_T^{(2)}
&=\frac{1}{R_{\rm se}}\Big[ S_{\rm sd,sd}R_{\rm sd}^2
+S_{\rm op,op}R_{\rm op}^2 +S_{\rm sd,op}R_{\rm sd}R_{\rm op} \nonumber\\
&+S_{\rm B,B}(\gamma_e B_z)^2 +\gamma_e B_z \left( S_{\rm B,sd}R_{\rm sd} +S_{\rm B,op}R_{\rm op} \right)  \Big].
\label{eq:appendix_second_order}
\end{align}
The dimensionless coefficients $S_{\alpha\beta}(p)$, listed in Table~\ref{tab:second_order_transverse}, are labeled by the pair of perturbations from which they originate, $\alpha,\beta\in\{{\rm sd}, {\rm op},B\}$. For $\alpha\neq\beta$, both orderings of the two perturbations contribute: $S_{\rm sd,op}$, for example, collects the terms proportional to $\mathcal{A}^{(\rm{sd})}\mathcal{A}^{(\rm{op})}$ and to $\mathcal{A}^{(\rm{op})}\mathcal{A}^{(\rm{sd})}$. The coefficients multiplying $R_{\rm sd}^2$, $R_{\rm op}^2$, $R_{\rm sd}R_{\rm op}$, and $(\gamma_e B_z)^2$ are real and modify the transverse relaxation rate. By contrast, the $S_{\rm B,sd}$ and $S_{\rm B,op}$ coefficients are purely imaginary and give second-order corrections to the spin precession frequency. The field-quadratic coefficient is $S_{\rm B,B}(p)=-C_T(p)$, so that its contribution to the transverse eigenvalue is $-C_T(p)(\gamma_e B_z)^2/R_{\rm se}$, corresponding to the positive residual spin-exchange broadening $C_T(p)(\gamma_e B_z)^2/R_{\rm se}$ derived in the main text. Closed-form expressions for $I=3/2$, $I=5/2$, and $I=7/2$ are given in Table~\ref{tab:CT}, with the limiting values at zero and full polarization providing useful checks.

For continuous circularly polarized D\textsubscript{1} pumping, the stationary polarization satisfies $\Rop=p\,(\Rop+\Rsd)$ and $\Rsd=(1-p)(\Rop+\Rsd)$, and the coefficients of Table~\ref{tab:second_order_transverse} satisfy the identities
\begin{align}
C_T(p)&=S_{\rm sd,sd}\,(1-p)^2+S_{\rm op,op}\,p^2+S_{\rm sd,op}\,p(1-p), \label{eq:identity_rates}\\
C_T(p)&= \frac{1}{2i} \left(S_{\rm B,sd}\,(1-p)+S_{\rm B,op}\,p \right), \label{eq:identity_mixed}
\end{align}
for all three nuclear spins. These identities reduce Eq.~\eqref{eq:appendix_second_order} at the pumped stationary state to the compact form of Eq.~\eqref{eq:lambdaT_zeta} in the main text. 

\begin{table*}[t]
\caption{ Second-order coefficients entering the slow transverse eigenvalue, Eq.~\eqref{eq:appendix_second_order}, for circularly polarized D\textsubscript{1} pumping ($s_{\rm ph}=1$). The coefficient $S_{\rm B,B}(p)=-C_T(p)$ is given separately in Table~\ref{tab:CT}. The last two columns give the limiting values at zero and full spin polarization.}
\label{tab:second_order_transverse}
\begin{ruledtabular}
\begin{tabular}{ccccc}
$I$ & Coefficient & Expression & $p=0$ & $p=1$ \\[2pt]
\hline
$\dfrac{3}{2}$ & $S_{\rm sd,sd}(p)$ & $\displaystyle \frac{(1+p^2)\mathcal P_{\rm sd,sd}^{(3/2)}(p)}{\mathcal D_{3/2}(p)}$ & $\displaystyle \frac{5}{108}$ & $\displaystyle \frac{9}{64}$ \\[9pt]
& $S_{\rm op,op}(p)$ & $\displaystyle \frac{(p-1)(1+p^2)\mathcal P_{\rm op,op}^{(3/2)}(p)}{\mathcal D_{3/2}(p)}$ & $\displaystyle \frac{5}{108}$ & $0$ \\[9pt]
& $S_{\rm sd,op}(p)$ & $\displaystyle \frac{(1+p^2)\mathcal P_{\rm sd,op}^{(3/2)}(p)}{\mathcal D_{3/2}(p)}$ & $\displaystyle \frac{5}{54}$ & $\displaystyle \frac{9}{64}$ \\[9pt]
& $S_{\rm B,sd}(p)$ & $\displaystyle \mathrm{i}\,\frac{(1+p^2)\mathcal P_{\rm B,sd}^{(3/2)}(p)}{\mathcal D_{3/2}(p)}$ & $\displaystyle \frac{5}{54}\,\mathrm{i}$ & $\displaystyle \frac{9}{64}\,\mathrm{i}$ \\[9pt]
& $S_{\rm B,op}(p)$ & $\displaystyle \mathrm{i}\,\frac{(p-1)(1+p^2)\mathcal P_{\rm B,op}^{(3/2)}(p)}{\mathcal D_{3/2}(p)}$ & $\displaystyle \frac{5}{54}\,\mathrm{i}$ & $0$ \\[9pt]
\hline
$\dfrac{5}{2}$ & $S_{\rm sd,sd}(p)$ & $\displaystyle \frac{2\mathcal A_{5/2}(p)\mathcal P_{\rm sd,sd}^{(5/2)}(p)}{\mathcal D_{5/2}(p)}$ & $\displaystyle \frac{210}{6859}$ & $\displaystyle \frac{5}{42}$ \\[9pt]
& $S_{\rm op,op}(p)$ & $\displaystyle \frac{2(p-1)\mathcal A_{5/2}(p)\mathcal P_{\rm op,op}^{(5/2)}(p)}{\mathcal D_{5/2}(p)}$ & $\displaystyle \frac{210}{6859}$ & $0$ \\[9pt]
& $S_{\rm sd,op}(p)$ & $\displaystyle \frac{2\mathcal A_{5/2}(p)\mathcal P_{\rm sd,op}^{(5/2)}(p)}{\mathcal D_{5/2}(p)}$ & $\displaystyle \frac{420}{6859}$ & $\displaystyle \frac{5}{42}$ \\[9pt]
& $S_{\rm B,sd}(p)$ & $\displaystyle \mathrm{i}\,\frac{2\mathcal A_{5/2}(p)\mathcal P_{\rm B,sd}^{(5/2)}(p)}{\mathcal D_{5/2}(p)}$ & $\displaystyle \frac{420}{6859}\,\mathrm{i}$ & $\displaystyle \frac{5}{42}\,\mathrm{i}$ \\[9pt]
& $S_{\rm B,op}(p)$ & $\displaystyle \mathrm{i}\,\frac{2(p-1)\mathcal A_{5/2}(p)\mathcal P_{\rm B,op}^{(5/2)}(p)}{\mathcal D_{5/2}(p)}$ & $\displaystyle \frac{420}{6859}\,\mathrm{i}$ & $0$ \\[9pt]
\hline
$\dfrac{7}{2}$ & $S_{\rm sd,sd}(p)$ & $\displaystyle \frac{\mathcal A_{7/2}(p)\mathcal P_{\rm sd,sd}^{(7/2)}(p)}{\mathcal D_{7/2}(p)}$ & $\displaystyle \frac{105}{5324}$ & $\displaystyle \frac{63}{640}$ \\[9pt]
& $S_{\rm op,op}(p)$ & $\displaystyle \frac{(p-1)\mathcal A_{7/2}(p)\mathcal P_{\rm op,op}^{(7/2)}(p)}{\mathcal D_{7/2}(p)}$ & $\displaystyle \frac{105}{5324}$ & $0$ \\[9pt]
& $S_{\rm sd,op}(p)$ & $\displaystyle \frac{\mathcal A_{7/2}(p)\mathcal P_{\rm sd,op}^{(7/2)}(p)}{\mathcal D_{7/2}(p)}$ & $\displaystyle \frac{105}{2662}$ & $\displaystyle \frac{63}{640}$ \\[9pt]
& $S_{\rm B,sd}(p)$ & $\displaystyle \mathrm{i}\,\frac{\mathcal A_{7/2}(p)\mathcal P_{\rm B,sd}^{(7/2)}(p)}{\mathcal D_{7/2}(p)}$ & $\displaystyle \frac{105}{2662}\,\mathrm{i}$ & $\displaystyle \frac{63}{640}\,\mathrm{i}$ \\[9pt]
& $S_{\rm B,op}(p)$ & $\displaystyle \mathrm{i}\,\frac{(p-1)\mathcal A_{7/2}(p)\mathcal P_{\rm B,op}^{(7/2)}(p)}{\mathcal D_{7/2}(p)}$ & $\displaystyle \frac{105}{2662}\,\mathrm{i}$ & $0$ \\[9pt]
\end{tabular}
\end{ruledtabular}
\end{table*}

For $I=3/2$ we define:
\begin{align}
\mathcal D_{3/2}(p) ={}& 4(3+p^2)^3(9+7p^2), \nonumber \\
\mathcal P_{\rm{sd,sd}}^{(3/2)}(p) ={}& 45+242p^2+p^4, \nonumber \\
\mathcal P_{\rm{op,op}}^{(3/2)}(p) ={}& -45+198p-44p^2+10p^3+9p^4, \nonumber \\
\mathcal P_{\rm{sd,op}}^{(3/2)}(p) ={}& 90-243p+484p^2-54p^3+2p^4+9p^5, \nonumber \\
\mathcal P_{\rm{B,sd}}^{(3/2)}(p) ={}& 90+241p^2-52p^4+9p^6, \nonumber \\
\mathcal P_{\rm{B,op}}^{(3/2)}(p) ={}& -90+153p-88p^2-34p^3+18p^4+9p^5. \nonumber
\end{align}
For $I=5/2$ we define:
\begin{align}
\mathcal A_{5/2}(p) ={}& (3+p^2)(1+3p^2), \nonumber \\
\mathcal D_{5/2}(p) ={}& (19+26p^2+3p^4)^3(100+269p^2+72p^4), \nonumber \\
\mathcal P_{\rm{sd,sd}}^{(5/2)}(p) ={}& 3500+50015p^2+64934p^4+44040p^6 \nonumber\\ &+19398p^8-447p^{10}, \nonumber \\
\mathcal P_{\rm{op,op}}^{(5/2)}(p) ={}& -3500+36400p-13615p^2+57116p^3 \nonumber\\ &-7818p^4+37656p^5-6384p^6+18252p^7 \nonumber\\ &-1146p^8-240p^9+207p^{10}, \nonumber \\
\mathcal P_{\rm{sd,op}}^{(5/2)}(p) ={}& 7000-39900p+100030p^2-70731p^3 \nonumber\\ &+129868p^4-45474p^5+88080p^6-24636p^7 \nonumber\\ &+38796p^8-906p^9-894p^{10}+207p^{11}, \nonumber \\
\mathcal P_{\rm{B,sd}}^{(5/2)}(p) ={}& 7000+60130p^2+59137p^4+42606p^6 \nonumber\\ &+14160p^8-1800p^{10}+207p^{12}, \nonumber \\
\mathcal P_{\rm{B,op}}^{(5/2)}(p) ={}& -7000+32900p-27230p^2+43501p^3 \nonumber\\ &-15636p^4+29838p^5-12768p^6+11868p^7 \nonumber\\ &-2292p^8-1386p^9+414p^{10}+207p^{11}. \nonumber
\end{align}
For $I=7/2$ we define:
\begin{align}
\mathcal A_{7/2}(p) &= (1+p^2)(1+6p^2+p^4), \nonumber \\
\mathcal D_{7/2}(p) &= 4(11+35p^2+17p^4+p^6)^3 \nonumber\\ &\times (99225+571041p^2+515203p^4+68931p^6), \nonumber \\
\mathcal P_{\rm sd,sd}^{(7/2)}(p) {}&= 10418625+291252780p^2+1182443136p^4 \nonumber\\ &+2309801388p^6+2621420442p^8+1262560676p^{10} \nonumber\\ &+358988936p^{12}+57011972p^{14}-1512675p^{16}, \nonumber \\
\mathcal P_{\rm op,op}^{(7/2)}(p) ={}& -10418625+195870150p-95382630p^2 \nonumber\\ &+955774134p^3-226669002p^4+1958259774p^5 \nonumber\\ &-351541614p^6+2351350734p^7-270069708p^8 \nonumber\\ &+1181731218p^9-80829458p^{10}+335207458p^{11} \nonumber\\ &-23781478p^{12}+55724522p^{13}-1287450p^{14} \nonumber\\ &-1249830p^{15}+262845p^{16}, \nonumber
\end{align}
\begin{align}
\mathcal P_{\rm sd,op}^{(7/2)}(p) {}&= 20837250-206288775p+582505560p^2 \nonumber\\ &-1051156764p^3+2364886272p^4-2184928776p^5 \nonumber\\ &+4619602776p^6-2702892348p^7+5242840884p^8 \nonumber\\ &-1451800926p^9+2525121352p^{10}-416036916p^{11} \nonumber\\ &+717977872p^{12}-79506000p^{13}+114023944p^{14} \nonumber\\ &-37620p^{15}-3025350p^{16}+262845p^{17}, \nonumber
\end{align}
\begin{align}
\mathcal P_{\rm B,sd}^{(7/2)}(p) {}&= 20837250+376216785p^2+1313729508p^4 \nonumber\\ &+2434674000p^6+2539948536p^8+1073320426p^{10} \nonumber\\ &+301940956p^{12}+34517944p^{14}-3062970p^{16} \nonumber\\ &+262845p^{18}, \nonumber
\end{align}
\begin{align}
\mathcal P_{\rm B,op}^{(7/2)}(p) {}&= -20837250+185451525p-190765260p^2 \nonumber\\ &+860391504p^3-453338004p^4+1731590772p^5 \nonumber\\ &-703083228p^6+1999809120p^7-540139416p^8 \nonumber\\ &+911661510p^9-161658916p^{10}+254378000p^{11} \nonumber\\ &-47562956p^{12}+31943044p^{13}-2574900p^{14} \nonumber\\ &-2537280p^{15}+525690p^{16}+262845p^{17}. \nonumber
\end{align}

\section{Numerical simulation of the density-matrix equation}
\label{sec:ApndxC: DM simulation}

The simulation solves the single-atom density-matrix equation including hyperfine evolution, the magnetic field, spin exchange, spin destruction, and high-buffer-gas-pressure optical pumping on the D\textsubscript{1} line,
\begin{equation}
\begin{split}
\frac{d\rho}{dt} &=-i A_{\rm hfs}[\mathbf I\cdot\mathbf S,\rho]
-i\,\gamma_e[\mathbf S\cdot\mathbf B,\rho]
+\Rse\left(\rho_{\langle \mathbf{S}\rangle}-\rho\right)\\
&+\Rop({\rho}_{\mathbf s_{\rm ph}/2}-\rho)+\Rsd ({\rho}_{\mathbf{0}}-\rho),
\label{eq:density_matrix}
\end{split}
\end{equation}
where $A_{\rm hfs}$ is the ground-state magnetic-dipole hyperfine coupling constant, and the mean-field state produced by each dynamical process is $\rho_{\bm{\mathcal{Q}}} =\left(\mathbb{1}_S/2 +2\bm{\mathcal{Q}}\cdot\!\mathbf  S\right)\otimes {\rm Tr}_{S}[\rho]$, with $\langle\mathbf S\rangle={\rm Tr}(\rho\,\mathbf S)$. Here $\mathbb{1}_S$ is the electron-spin identity and ${\rm Tr}_S$ denotes the trace over the electron spin. The pump beam and the dc magnetic field are directed along $z$, and circularly polarized pumping, $\mathbf s_{\rm ph}=\hat{\mathbf z}$, is used throughout the numerical calculations. The simulation is performed for \textsuperscript{87}Rb ($I=3/2$), and the rapidly oscillating hyperfine coherences are projected out \cite{Savukov2005}.

Writing the equation of motion as $\dot\rho=\mathcal{L}(\rho)$, where $\mathcal{L}$ acts nonlinearly on $\rho$ through the spin-exchange mean field, the stationary state $\rho_{\rm ss}$ is obtained by propagating Eq.~\eqref{eq:density_matrix} from the unpolarized state $\rho(0)=\mathbb{1}/[2(2I+1)]$ until convergence. For the sudden pumping and relaxation processes considered here, the stationary state is of spin-temperature form \cite{Appelt1998}, as we confirm numerically, with polarization $p=2|\langle\mathbf S\rangle_{\rm ss}|$.

The dynamics of small deviations $\rho=\rho_{\rm ss}+\delta\rho$ are governed by the linearized Liouvillian $\mathcal{L}_{\rm lin}$ introduced in Eq.~\eqref{eq:linearized}, that is, by the Jacobian of the nonlinear dynamics evaluated at the stationary state. In Liouville space, the density matrix is arranged into a column vector with components $\rho_j$, and the right-hand side of Eq.~\eqref{eq:density_matrix} becomes a vector-valued function with components $\mathcal{L}_i(\rho)$. The linearized Liouvillian is then the matrix of partial derivatives evaluated at the stationary state,
\begin{equation}
\left[\mathcal{L}_{\rm lin}\right]_{ij}
=
\left.
\frac{\partial \mathcal{L}_i(\rho)}{\partial \rho_j}
\right|_{\rho=\rho_{\rm ss}} .
\end{equation}
Because the spin-exchange mean-field state depends on the instantaneous electron spin, the linearization retains the variation $\delta\langle\mathbf S\rangle={\rm Tr}(\mathbf S\,\delta\rho)$ in addition to the stationary value $\langle\mathbf S\rangle_{\rm ss}$. To first order in $\delta\rho$,
\begin{equation}
\delta\rho_{\langle\mathbf S\rangle}
=\left(\frac{\mathbb 1_S}{2}
+2\langle\mathbf S\rangle_{\rm ss}\cdot\mathbf S\right)
\otimes {\rm Tr}_S[\delta\rho]
+2\,\delta\langle\mathbf S\rangle\cdot\mathbf S
\otimes {\rm Tr}_S[\rho_{\rm ss}].
\label{eq:SE_feedback}
\end{equation}

For $\mathbf{B}=B_z\hat{\mathbf z}$, the longitudinal and transverse blocks of $\mathcal{L}_{\rm lin}$ remain decoupled. Physical deviations $\delta\rho$ are traceless, since $\rho$ and $\rho_{\rm ss}$ are both normalized, and the eigenmodes of $\mathcal{L}_{\rm lin}$ respect this automatically. Because the dynamics preserve the trace, $\Tr(\mathcal{L}_{\rm lin}\,\delta\rho)=0$ for any $\delta\rho$, so a right eigenmode satisfying $\mathcal{L}_{\rm lin}r_n=\lambda_n r_n$ obeys $\lambda_n \Tr (r_n)=0$. Every eigenmode with nonzero eigenvalue is therefore traceless, and the only mode with nonzero trace is a zero-eigenvalue mode that reflects the freedom in the normalization of $\rho$. This mode carries no physical excitation and is discarded. Numerically, we retain the modes with $\left|\Tr (r_n)\right|<\epsilon_{\rm tr}$ and use a small eigenvalue tolerance to remove numerical null modes.

The physical longitudinal and transverse relaxation modes are then identified by their overlap with the electronic-spin observables. After projecting out the fast hyperfine coherences, each right eigenmode is normalized to unit Frobenius norm, $\|r_n\|_{\rm F}=1$, and assigned the longitudinal and transverse weights
\begin{equation}
w_{z,n}
=
\frac{\left|\Tr(S_z r_n)\right|}{\|S_z\|_{\rm F}},
\qquad
w_{+,n}
=
\frac{\left|\Tr(S_+ r_n)\right|}{\|S_+\|_{\rm F}},
\end{equation}
with $S_+=S_x+iS_y$, which measure the visibility of the mode in $\langle S_z\rangle$ and $\langle S_+\rangle$, respectively. The slow longitudinal mode is the mode with appreciable longitudinal weight and the smallest decay rate $-\operatorname{Re}\lambda_n$, and the transverse modes are selected analogously by their transverse weight. A weight is considered appreciable if it exceeds $10^{-3}$ of the largest corresponding observable weight among the modes retained in the physical subspace. At finite $B_z$, the two slowest transverse modes are the complex-conjugate pair $\lambda_T$ and $\lambda_T^*$ of Sec.~\ref{sec:theory}.

The results shown in Fig.~\ref{fig:DM_ver} are obtained with $\Rse=10^{5}~\si{\per\second}$ and $\Rsd=10^{2}~\si{\per\second}$, the polarization being scanned through the optical-pumping rate according to $\Rop=p\,\Rsd/(1-p)$. Panel~(b) uses longitudinal fields $B_z=0$, $100$, and $300~\si{\nano\tesla}$.

\section*{Acknowledgment}
We thank Michael Romalis for useful discussions.

\bibliography{refs}

@article{ShahRomalis,
  title = {Spin-exchange relaxation-free magnetometry using elliptically polarized light},
  author = {Shah, V. and Romalis, M. V.},
  journal = {Phys. Rev. A},
  volume = {80},
  issue = {1},
  pages = {013416},
  numpages = {6},
  year = {2009},
  month = {Jul},
  publisher = {American Physical Society},
  doi = {10.1103/PhysRevA.80.013416},
  url = {https://link.aps.org/doi/10.1103/PhysRevA.80.013416}
}

@article{Ledbetter2008SERF,
  title = {Spin-exchange-relaxation-free magnetometry with {Cs} vapor},
  author = {Ledbetter, M. P. and Savukov, I. M. and Acosta, V. M. and Budker, D. and Romalis, M. V.},
  journal = {Phys. Rev. A},
  volume = {77},
  issue = {3},
  pages = {033408},
  numpages = {7},
  year = {2008},
  month = {Mar},
  publisher = {American Physical Society},
  doi = {10.1103/PhysRevA.77.033408},
  url = {https://link.aps.org/doi/10.1103/PhysRevA.77.033408}
}

@article{HapperReview1972,
  title = {Optical Pumping},
  author = {Happer, William},
  journal = {Rev. Mod. Phys.},
  volume = {44},
  issue = {2},
  pages = {169--249},
  numpages = {0},
  year = {1972},
  month = {Apr},
  publisher = {American Physical Society},
  doi = {10.1103/RevModPhys.44.169},
  url = {https://link.aps.org/doi/10.1103/RevModPhys.44.169}
}

@article{Appelt1998,
  title = {Theory of spin-exchange optical pumping of ${}^{3}\mathrm{He}$ and ${}^{129}\mathrm{Xe}$},
  author = {Appelt, S. and Baranga, A. Ben-Amar and Erickson, C. J. and Romalis, M. V. and Young, A. R. and Happer, W.},
  journal = {Phys. Rev. A},
  volume = {58},
  issue = {2},
  pages = {1412--1439},
  numpages = {0},
  year = {1998},
  month = {Aug},
  publisher = {American Physical Society},
  doi = {10.1103/PhysRevA.58.1412},
  url = {https://link.aps.org/doi/10.1103/PhysRevA.58.1412}
}

@article{MouloudakisVasilakis2026,
  title = {Quantum sensitivity limits and dynamical correlations in alkali-vapor sensors},
  author = {Mouloudakis, K. and Koutrouli, V. and Kominis, I.K. and Mitchell, M.W. and Vasilakis, G.},
  journal = {Phys. Rev. Appl.},
  volume = {25},
  issue = {2},
  pages = {024009},
  numpages = {18},
  year = {2026},
  month = {Feb},
  publisher = {American Physical Society},
  doi = {10.1103/mbdq-md8f},
  url = {https://link.aps.org/doi/10.1103/mbdq-md8f}
}

@article{Tang2025,
  title = {Magnetic-resonance linewidth of alkali-metal vapor in the unresolved {Zeeman}-resonance regime},
  author = {Tang, Feng and Zhao, Nan},
  journal = {Phys. Rev. A},
  volume = {111},
  issue = {1},
  pages = {013103},
  numpages = {11},
  year = {2025},
  month = {Jan},
  publisher = {American Physical Society},
  doi = {10.1103/PhysRevA.111.013103},
  url = {https://link.aps.org/doi/10.1103/PhysRevA.111.013103}
}

@article{Kornack2005,
  title = {Nuclear Spin Gyroscope Based on an Atomic Comagnetometer},
  author = {Kornack, T. W. and Ghosh, R. K. and Romalis, M. V.},
  journal = {Phys. Rev. Lett.},
  volume = {95},
  issue = {23},
  pages = {230801},
  numpages = {4},
  year = {2005},
  month = {Nov},
  publisher = {American Physical Society},
  doi = {10.1103/PhysRevLett.95.230801},
  url = {https://link.aps.org/doi/10.1103/PhysRevLett.95.230801}
}

@article{Budker2007,
  title = {Optical magnetometry},
  volume = {3},
  ISSN = {1745-2481},
  url = {http://dx.doi.org/10.1038/nphys566},
  DOI = {10.1038/nphys566},
  number = {4},
  journal = {Nature Physics},
  publisher = {Springer Science and Business Media LLC},
  author = {Budker,  Dmitry and Romalis,  Michael},
  year = {2007},
  month = Apr,
  pages = {227–234}
}

@article{Kominis2003,
  title = {A subfemtotesla multichannel atomic magnetometer},
  volume = {422},
  ISSN = {1476-4687},
  url = {http://dx.doi.org/10.1038/nature01484},
  DOI = {10.1038/nature01484},
  number = {6932},
  journal = {Nature},
  publisher = {Springer Science and Business Media LLC},
  author = {Kominis,  I. K. and Kornack,  T. W. and Allred,  J. C. and Romalis,  M. V.},
  year = {2003},
  month = Apr,
  pages = {596–599}
}

@article{wqqq-s2bz,
  title = {$^{3}\mathrm{He}\text{\ensuremath{-}}^{21}\mathrm{Ne}$ {Ramsey} Comagnetometer with Sub-n{H}z Frequency Resolution},
  author = {Zhang, Shaobo and Wang, Jingyao and Sun, George and van de Wetering, Johannes J. and Romalis, Michael V.},
  journal = {Phys. Rev. Lett.},
  volume = {136},
  issue = {20},
  pages = {203201},
  numpages = {6},
  year = {2026},
  month = {May},
  publisher = {American Physical Society},
  doi = {10.1103/wqqq-s2bz},
  url = {https://link.aps.org/doi/10.1103/wqqq-s2bz}
}

@article{Vasilakis2009,
  title = {Limits on New Long Range Nuclear Spin-Dependent Forces Set with a $\mathbf{K}\mathrm{\text{\ensuremath{-}}}^{3}\mathrm{He}$ Comagnetometer},
  author = {Vasilakis, G. and Brown, J. M. and Kornack, T. W. and Romalis, M. V.},
  journal = {Phys. Rev. Lett.},
  volume = {103},
  issue = {26},
  pages = {261801},
  numpages = {4},
  year = {2009},
  month = {Dec},
  publisher = {American Physical Society},
  doi = {10.1103/PhysRevLett.103.261801},
  url = {https://link.aps.org/doi/10.1103/PhysRevLett.103.261801}
}

@article{PhysRevApplied.19.044092,
  title = {Optimization of Nuclear Polarization in an Alkali-Noble Gas Comagnetometer},
  author = {Klinger, Emmanuel and Liu, Tianhao and Padniuk, Mikhail and Engler, Martin and Kornack, Thomas and Pustelny, Szymon and Jackson Kimball, Derek F. and Budker, Dmitry and Wickenbrock, Arne},
  journal = {Phys. Rev. Appl.},
  volume = {19},
  issue = {4},
  pages = {044092},
  numpages = {11},
  year = {2023},
  month = {Apr},
  publisher = {American Physical Society},
  doi = {10.1103/PhysRevApplied.19.044092},
  url = {https://link.aps.org/doi/10.1103/PhysRevApplied.19.044092}
}

@article{HapperTang1973,
  title   = {Spin-Exchange Shift and Narrowing of Magnetic Resonance Lines in Optically Pumped Alkali Vapors},
  author  = {Happer, W. and Tang, H.},
  journal = {Phys. Rev. Lett.},
  volume  = {31},
  pages   = {273},
  year    = {1973},
  doi     = {10.1103/PhysRevLett.31.273}
}

@article{HapperTam1977,
  title   = {Effect of rapid spin exchange on the magnetic-resonance spectrum of alkali vapors},
  author  = {Happer, W. and Tam, A. C.},
  journal = {Phys. Rev. A},
  volume  = {16},
  pages   = {1877},
  year    = {1977},
  doi     = {10.1103/PhysRevA.16.1877}
}

@article{Allred2002,
  title   = {High-Sensitivity Atomic Magnetometer Unaffected by Spin-Exchange Relaxation},
  author  = {Allred, J. C. and Lyman, R. N. and Kornack, T. W. and Romalis, M. V.},
  journal = {Phys. Rev. Lett.},
  volume  = {89},
  pages   = {130801},
  year    = {2002},
  doi     = {10.1103/PhysRevLett.89.130801}
}

@article{Savukov2005,
  title   = {Effects of spin-exchange collisions in a high-density alkali-metal vapor in low magnetic fields},
  author  = {Savukov, I. M. and Romalis, M. V.},
  journal = {Phys. Rev. A},
  volume  = {71},
  pages   = {023405},
  year    = {2005},
  doi     = {10.1103/PhysRevA.71.023405}
}

@article{KatzFirstenberg2018,
  title   = {Synchronization of strongly interacting alkali-metal spins},
  author  = {Katz, O. and Firstenberg, O.},
  journal = {Phys. Rev. A},
  volume  = {98},
  pages   = {012712},
  year    = {2018},
  doi     = {10.1103/PhysRevA.98.012712}
}

@article{Xiao2021,
  title   = {Atomic spin-exchange collisions in magnetic fields},
  author  = {Xiao, W. and Wu, T. and Peng, X. and Guo, H.},
  journal = {Phys. Rev. A},
  volume  = {103},
  pages   = {043116},
  year    = {2021},
  doi     = {10.1103/PhysRevA.103.043116}
}

@article{Dikopoltsev_2025,
  title = {Suppressing the Decoherence of Alkali-Metal Spins at Low Magnetic Fields},
  author = {Dikopoltsev, Mark and Berrebi, Avraham and Levy, Uriel and Katz, Or},
  journal = {Phys. Rev. Lett.},
  volume = {134},
  issue = {14},
  pages = {143201},
  numpages = {6},
  year = {2025},
  month = {Apr},
  publisher = {American Physical Society},
  doi = {10.1103/PhysRevLett.134.143201},
  url = {https://link.aps.org/doi/10.1103/PhysRevLett.134.143201}
}

@article{Padniuk2022,
 author={Padniuk, Mikhail
and Kopciuch, Marek
and Cipolletti, Riccardo
and Wickenbrock, Arne
and Budker, Dmitry
and Pustelny, Szymon},
title={Response of atomic spin-based sensors to magnetic and nonmagnetic perturbations},
journal={Scientific Reports},
year={2022},
month={Jan},
day={10},
volume={12},
number={1},
pages={324},
issn={2045-2322},
doi={10.1038/s41598-021-03609-w},
url={https://doi.org/10.1038/s41598-021-03609-w}
}

@article{PhysRevApplied.21.014023,
  title = {All-optical dual-axis zero-field atomic magnetometer using light-shift modulation},
  author = {Li, Xiaoyu and Han, Bangcheng and Zhang, Kaixuan and Liu, Ziao and Wang, Shuying and Yan, Yifan and Lu, Jixi},
  journal = {Phys. Rev. Appl.},
  volume = {21},
  issue = {1},
  pages = {014023},
  numpages = {7},
  year = {2024},
  month = {Jan},
  publisher = {American Physical Society},
  doi = {10.1103/PhysRevApplied.21.014023},
  url = {https://link.aps.org/doi/10.1103/PhysRevApplied.21.014023}
}

@article{PhysRevA.107.043110,
  title = {Partial measurements of the total field gradient and the field-gradient tensor using an atomic magnetic gradiometer},
  author = {Yu, Q.-Q. and Liu, S.-Q. and Wang, X.-K. and Sheng, D.},
  journal = {Phys. Rev. A},
  volume = {107},
  issue = {4},
  pages = {043110},
  numpages = {7},
  year = {2023},
  month = {Apr},
  publisher = {American Physical Society},
  doi = {10.1103/PhysRevA.107.043110},
  url = {https://link.aps.org/doi/10.1103/PhysRevA.107.043110}
}

@article{Hedges2025,
  title = {Dual axis atomic magnetometer and gyroscope enabled by nuclear spin perturbation},
  volume = {27},
  ISSN = {1367-2630},
  url = {http://dx.doi.org/10.1088/1367-2630/adc6b3},
  DOI = {10.1088/1367-2630/adc6b3},
  number = {4},
  journal = {New Journal of Physics},
  publisher = {IOP Publishing},
  author = {Hedges,  Morgan and Papneja,  Ankit and Paul,  Karun and Buchler,  Ben C},
  year = {2025},
  month = Apr,
  pages = {043016}
}

@article{Terrano2022,
doi = {10.1088/2058-9565/ac1ae0},
url = {https://doi.org/10.1088/2058-9565/ac1ae0},
year = {2021},
month = {nov},
publisher = {IOP Publishing},
volume = {7},
number = {1},
pages = {014001},
author = {Terrano, W A and Romalis, M V},
title = {Comagnetometer probes of dark matter and new physics},
journal = {Quantum Science and Technology},
}

@article{TransientOE2021,
author = {Junjian Tang and Yan Yin and YueYang Zhai and Binquan Zhou and Bangcheng Han and Hongying Yang and Gang Liu},
journal = {Opt. Express},
number = {6},
pages = {8333--8343},
publisher = {Optica Publishing Group},
title = {Transient dynamics of atomic spin in the spin-exchange-relaxation-free regime},
volume = {29},
month = {Mar},
year = {2021},
url = {https://opg.optica.org/oe/abstract.cfm?URI=oe-29-6-8333},
doi = {10.1364/OE.418776},
}

@article{PhysRevA.109.L040802,
  title = {Anomalous noise spectra in a spin-exchange-relaxation-free alkali-metal vapor},
  author = {Mouloudakis, K. and Kong, J. and Sierant, A. and Arkin, E. and Hern\'andez Ruiz, M. and Jim\'enez-Mart\'{\i}nez, R. and Mitchell, M. W.},
  journal = {Phys. Rev. A},
  volume = {109},
  issue = {4},
  pages = {L040802},
  numpages = {5},
  year = {2024},
  month = {Apr},
  publisher = {American Physical Society},
  doi = {10.1103/PhysRevA.109.L040802},
  url = {https://link.aps.org/doi/10.1103/PhysRevA.109.L040802}
}

\end{document}